\documentclass[conference]{IEEEtran}
\pdfoutput=1
\usepackage{cite}
\usepackage{amsmath,amssymb,amsfonts}
\usepackage{algorithmic}
\usepackage{graphicx}
\usepackage{textcomp}
\usepackage{multirow}
\usepackage{arydshln}
\usepackage{url}

\usepackage{bm}
\usepackage{makecell}
\usepackage{booktabs}
\usepackage[linesnumbered,ruled,vlined]{algorithm2e}
\usepackage[table]{xcolor}

\usepackage{minted}

\SetKwFunction{ApplyTransform}{ApplyTransform}
\SetKwProg{Fn}{Function}{:}{}

\usepackage[most]{tcolorbox}
\usetikzlibrary{shadows}  
\tcbset{
  RQConBox/.style={
    enhanced,
    frame hidden,
    boxrule=0pt,
    arc=3pt, outer arc=3pt,
    colback=white, colbacklower=blue!3!white,
    left=8pt, right=6pt, top=4pt, bottom=4pt,
    borderline west={3pt}{0pt}{blue!60!black},
    drop shadow={shadow xshift=0.5pt, shadow yshift=-0.5pt, opacity=0.15},
    fontupper=\normalsize\rmfamily,
  }
}

\def\BibTeX{{\rm B\kern-.05em{\sc i\kern-.025em b}\kern-.08em
    T\kern-.1667em\lower.7ex\hbox{E}\kern-.125emX}}

\definecolor{rqbg}{HTML}{E6F0FA} 

\newlength{\IEEEsavedparindent}
\AtBeginDocument{\setlength{\IEEEsavedparindent}{\parindent}}

\tcbset{
  conclbox/.style={
    enhanced,
    breakable,                    
    colback=rqbg,                 
    boxrule=0pt,
    arc=0pt,
    left=0pt, right=0pt, top=0pt, bottom=0pt,
    boxsep=0pt,
    before upper={
      \vspace{0.8mm}\strut
      \setlength{\parindent}{\IEEEsavedparindent}\indent
    },
    after upper={\strut},
    before skip=0pt,
    after skip=0pt,
  }
}

\newcommand{\conclbox}[1]{%
  \par
  \begin{tcolorbox}[conclbox]
    #1
  \end{tcolorbox}%
}

\tcbset{
  colback=blue!10,   
  colframe=white,    
  boxrule=0pt,       
  left=0pt,right=0pt,top=2pt,bottom=2pt,
  enhanced,
  breakable
}

\usepackage{fancyhdr}
\fancypagestyle{firstpage}{%
  \fancyhf{} 
  \fancyfoot[C]{This is the author’s version of the work accepted at ASE 2025. 
  The definitive version will be available at IEEE/ACM Digital Library.}
}

\begin{document}

\title{Effective Code Membership Inference for Code Completion Models via Adversarial Prompts\\
}

\author{
Yuan Jiang$^{\dagger}$, Zehao Li$^{\dagger}$, Shan Huang$^{\dagger}$, Christoph Treude$^{\S}$, Xiaohong Su$^{\dagger}$, Tiantian Wang$^{\dagger}$ \\
$^{\dagger}$Harbin Institute of Technology $^{\S}$Singapore Management University \\
\{jiangyuan, sxh, wangtiantian\}@hit.edu.cn, \{2021110768, 2022110145\}@stu.hit.edu.cn, ctreude@smu.edu.sg
}



\maketitle
\thispagestyle{firstpage} 

\begin{abstract}
Membership inference attacks (MIAs) on code completion models offer an effective way to assess privacy risks by inferring whether a given code snippet was part of the training data.
Existing black- and gray-box MIAs rely on expensive surrogate models or manually crafted heuristic rules, which limit their ability to capture the nuanced memorization patterns exhibited by over-parameterized code language models.
To address these challenges, we propose \textbf{AdvPrompt-MIA}, a method specifically designed for code completion models, combining code-specific adversarial perturbations with deep learning.
The core novelty of our method lies in designing a series of adversarial prompts that induce variations in the victim code model’s output. By comparing these outputs with the ground-truth completion, we construct feature vectors to train a classifier that automatically distinguishes member from non-member samples. This design allows our method to capture richer memorization patterns and accurately infer training set membership.
We conduct comprehensive evaluations on widely adopted models, such as Code Llama 7B, over the APPS and HumanEval benchmarks. The results show that our approach consistently outperforms state-of-the-art baselines, with AUC gains of up to 102\%.
In addition, our method exhibits strong transferability across different models and datasets, underscoring its practical utility and generalizability.
\end{abstract}

\begin{IEEEkeywords}
Code LLMs, Membership Inference Attacks, Adversarial Prompts, Robustness
\end{IEEEkeywords}

\section{Introduction}

Large language models (LLMs) have shown remarkable success in natural language processing by learning complex semantic and syntactic patterns from large-scale text corpora~\cite{devlin2019bert,raffel2020t5}. This success has extended to the domain of source code, where code-specific LLMs (code LLMs) trained on billions of lines of code~\cite{Feng2020} now support tasks such as code completion~\cite{wang2021codet5}, code summarization~\cite{ahmed2024automatic,zhu2024effectiveness}, and vulnerability detection~\cite{jiang2024stagedvulbert}, and are integrated into tools like GitHub Copilot~\cite{copilot2021github} and AWS CodeWhisperer~\cite{amazon2023codewhisperer}.

Despite their impressive capabilities, code LLMs remain vulnerable to a variety of security and privacy threats, including adversarial perturbations~\cite{liu2024eatvul}, data poisoning~\cite{schuster2021poisoning,cotroneo2024vulnerabilities}, and privacy leakage~\cite{niu2023codexleaks,yang2024unveiling,han2025codebreaker}. Among these, privacy leakage is particularly concerning due to its implications for sensitive information exposure and potential legal violations, often stemming from the memorization behavior of code LLMs~\cite{rabin2023memorization,carlini2022quantifying,huang2024your}. MIAs, which infer whether a specific code sample was included in the training set of a target code LLMs~\cite{yang2024survey}, provide a means to uncover and quantify such memorization, thereby helping mitigate this risk.

The practical value of successful MIAs on code LLMs is multifaceted. From a privacy perspective, leaked membership information may expose sensitive code snippets containing API keys, credentials, or personal identifiers~\cite{niu2023codexleaks}. MIAs can assess the extent of such leakage in code LLMs and help mitigate it by guiding the design of effective defenses. From a legal perspective, memorization and unauthorized reproduction of code from open-source repositories may violate license agreements (e.g., GPL, CC-BY-SA) if proper attribution is omitted~\cite{al2024traces,sun2022coprotector}. Employing MIAs can promote the use of policy-compliant training datasets. From an accountability perspective, membership leakage may serve as forensic evidence for identifying unauthorized training or data misuse. MIAs can thus support data governance and regulatory compliance~\cite{sun2023codemark}.


Therefore, advancing research on MIAs is essential for promoting responsible use of code LLMs and better mitigating the privacy risks associated with model memorization.
While membership inference techniques can be misused (e.g., to probe models for memorized credentials), our work is aimed exclusively at understanding and mitigating privacy risks in code models. The methods we propose are intended to support research, auditing, and regulatory compliance efforts.

Existing black- and gray-box MIAs for code models generally fall into three paradigms~\cite{yang2024survey}. The first trains a substitute model to approximate the target and performs shadow-model attacks based on its behavior~\cite{yang2024gotcha}. The second directly compares the model's generated output with the ground-truth completion, flagging exact matches or high semantic similarity as evidence of membership~\cite{alkaswan2024traces, yang2024unveiling}. The third analyzes differences in internal representations when the model is queried with syntactic variants of the same input (e.g., lowercase versus uppercase code), assuming that large representational shifts signal memorization~\cite{zhang2023code}.
However, due to their high computational overhead, fragile reliance on output similarity, and inability to capture richer memorization patterns, these strategies still leave considerable room for improvement in reliably detecting memorization in large-scale code LLMs.

In this study, we propose AdvPrompt-MIA, which infers code membership status by leveraging a series of semantics-preserving and code-specific adversarial prompts to expose model memorization.
The core insight is that code completion models tend to produce more stable outputs for training samples than for unseen inputs. This behavioral stability is particularly pronounced when the input is modified with small, functionality-preserving perturbations, such as inserting dead loops or renaming variables. That means, if the input is a memorized training sample, the model’s output remains largely consistent even after applying code perturbations, staying close to the original completion. In contrast, for non-member inputs, the same perturbations often lead to noticeably different outputs. This distinction between members and non-members under perturbations forms the basis of our approach, and similar patterns have also been observed in prior work on adversarial machine learning~\cite{tanay2016boundary,tian2018detecting,hu2019new,choquette2021label}.

AdvPrompt-MIA operationalizes this insight by introducing five types of semantics-preserving perturbations and quantifying the model’s behavioral differences in response to perturbed versus original inputs. These differences are measured using both similarity and perplexity metrics, from which we construct feature vectors that characterize the model’s output patterns. The resulting features are used to train a deep learning classifier that automatically learns discriminative signals indicative of membership. This design enables our method to detect subtle behavioral shifts and reliably infer whether a given input-output pair was part of the model’s training data.

We evaluate our method on the Code Llama 7B model~\cite{roziere2023code} and observe substantial AUC improvements ranging from 63.8\% to 102.0\% over state-of-the-art baselines on APPS and HumanEval. We further demonstrate that our approach generalizes well to other code models, including Deepseek-Coder 7B~\cite{guo2024deepseek}, StarCoder2 7B~\cite{lozhkov2024starcoder}, Phi-2 2.7B~\cite{javaheripi2023phi}, and WizardCoder 7B~\cite{luo2023wizardcoder}. Additionally, we conduct transferability experiments in which the MIA classifier is trained and tested across different models or datasets, showing that our method maintains strong performance under cross-model and cross-dataset settings.
The main contributions are as follows.

\begin{itemize}

    \item We propose a novel MIA framework, AdvPrompt-MIA, which leverages semantics-preserving adversarial prompts to perturb input code and employs a deep learning classifier to automatically learn discriminative patterns from the model's behavioral variations.

    \item We design five types of semantics-preserving perturbations, each crafted to elicit measurable differences in model behavior. Empirical results demonstrate that these perturbations effectively amplify membership signals and enhance inference accuracy.

    \item We conduct comprehensive experiments on multiple code LLMs and two widely used benchmarks (APPS and HumanEval), showing that our approach consistently outperforms state-of-the-art baselines across a range of models and evaluation settings.
\end{itemize}

\section{Motivation and Background}
\subsection{MIA Definition and Representative Attack Methods}
\label{sec:MIA_define}
MIAs aim to determine whether a given data sample was used during the training of a target machine learning model. In the context of code completion, each data point is represented as a pair $(x, y)$, where $x$ is a partial code snippet (i.e., the input prefix), and $y$ is its corresponding expected completion. Given black-box access to a trained code completion model $M$, the adversary can query the model with $x$ and obtain the generated output $\hat{y} = M(x)$. The objective is to infer whether the pair $(x, y)$ was present in the training dataset $D_{\text{in}}$ of $M$.
Formally, an MIA constructs a binary classifier $G$, which takes as input the triple $(x, y, \hat{y})$ and predicts a membership label:
\[
G(x, y, \hat{y}) \in \{0, 1\},
\]
where a prediction of 1 indicates that $(x, y)$ is a member sample (i.e., $ (x, y) \in D_{\text{in}} $), and 0 otherwise.

A commonly used baseline defines $G$ as a heuristic rule that compares the ground-truth completion $y$ with the model output $\hat{y}$. One typical example is GT-Match, which infers membership by checking whether $\hat{y}$ exactly matches $y$:

\[
G_{\text{exact}}(x, y, \hat{y}) = \mathbb{1}\left[ \hat{y} = y \right],
\]
where $\mathbb{1}[\cdot]$ denotes the indicator function. This method assumes that if the model reproduces $y$ exactly given $x$, it is likely that $(x, y)$ was memorized during training. However, this approach often suffers from low true positive rates, particularly for large code models that may generalize to produce semantically correct but non-identical completions. In such cases, even if $x$ was included in the training set, the model may generate $\hat{y} \neq y$~\cite{alkaswan2024traces}.

To address these limitations, many prior studies adopt a shadow model strategy~\cite{shokri2017membership}. In this setting, the adversary trains a local surrogate model $\hat{M}$ on data sampled from the same or a similar distribution as the target model's training set. Since the attacker has full control over $\hat{M}$, they can label any sample $(x, y)$ as a member or non-member based on whether it was in $\hat{M}$’s training data. The resulting dataset $\mathcal{D}_{\text{attack}}$ is then used to train the attack classifier $G$. Formally, the adversary constructs:
\[
\mathcal{D}_{\text{attack}} = \{(x_i, y_i, \hat{M}(x_i), m_i)\}_{i=1}^{n},
\]
where $m_i = 1$ if $(x_i, y_i)$ was used to train $\hat{M}$, and $m_i = 0$ otherwise.
This approach has been shown to be effective against small-scale code LLMs~\cite{yang2024survey}. However, its application to larger models remains limited. Training shadow models that approximate large commercial code models is computationally expensive, and small-scale surrogates often fail to capture the behavior of large models, thereby reducing the transferability and effectiveness of the attack~\cite{niu2023codexleaks}.

\subsection{Motivation for the Proposed Method and How It Differs from Prior Work}
\label{sec:our_intuition}
The methods described in Section~\ref{sec:MIA_define} share a common intuition: if a model has seen an input-output pair \( (x, y) \) during training, it is more likely to generate an output \( \hat{y} \) that closely matches \( y \). For example, GT-Match directly reflects this intuition by inferring membership based on exact string matching between \( \hat{y} \) and \( y \). Shadow-model–based approaches also follow this principle by learning to classify membership using feature representations derived from \( x \), \( y \), and \( \hat{y} \), implicitly modeling the alignment between the generated and reference outputs.
However, this intuition tends to be less pronounced for larger code LLMs. As these models are trained on massive and diverse datasets, they exhibit strong generalization and may generate completions that differ from the ground truth even for training samples. Conversely, for common functional code, they may produce outputs identical or similar to the ground truth despite having never seen the exact pair during training.

Our method is motivated by a different yet complementary observation: membership signals are more robust under semantics-preserving perturbations~\cite{choquette2021label}. Specifically, when a sample \( (x, y) \) is part of the training data, the model tends to generate completions that maintain a stable relationship with \( y \) across small, functionality-preserving changes to \( x \). This behavioral consistency reflects underlying memorization. In contrast, for non-member samples, the model's outputs are more sensitive to such perturbations, resulting in greater variation in their alignment with \( y \). This divergence under controlled perturbations offers a fine-grained signal for distinguishing members from non-members.

\begin{figure}[htbp]
\centering
\includegraphics[width=0.5\textwidth]{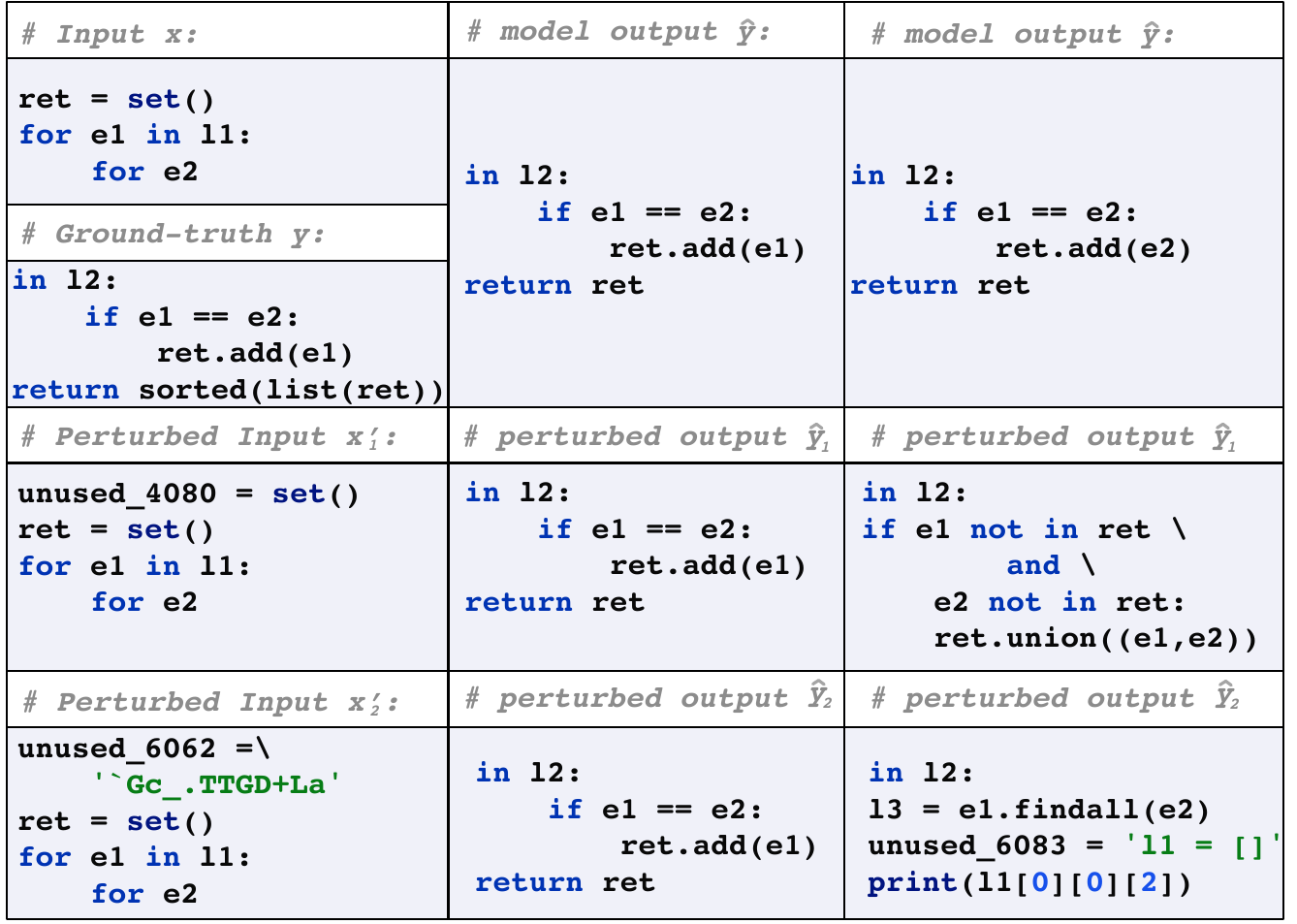}
\caption{An illustrative example showing the code pair $(x, y)$, along with the model’s original and perturbed outputs when $(x, y)$ is a memorized sample (middle column) and when it is a non-memorized sample (right column).}
\label{fig:motivation_example}
\end{figure}

To further clarify this intuition, Fig.~\ref{fig:motivation_example} presents a representative example illustrating the victim model's responses to an input \( x \) and its perturbed variants, conditioned on whether the pair \( (x, y) \) appears in the training set. As shown, the model tends to generate similar completions for the original input \( x \) in both member and non-member cases, which limits the effectiveness of GT-Match and shadow-model–based methods. In contrast, when semantics-preserving perturbations are applied, the outputs for non-member samples exhibit greater deviation from the ground truth \( y \) (right column), whereas member samples remain stable (middle column), indicating stronger memorization and reduced sensitivity to input changes.

Motivated by this example, we propose a deep learning-based method that captures both the consistency and variation in model responses to semantics-preserving perturbations, enabling more effective identification of subtle memorization patterns through aggregated observations.

\section{Overview of Our Attack Methodology}
This section defines the threat model in terms of the adversary’s goals, knowledge, and capabilities~\cite{apruzzese2023real}, and then briefly describes our method’s workflow under these assumptions.

\subsection{Threat Model}
\paragraph{Adversary’s Goal}
Let $M$ be a code completion model trained on a private dataset $D_{in} = \{(x_i, y_i)\}_{i=1}^{n}$, where each pair $(x_i, y_i)$ represents an input code snippet and its corresponding ground truth completion. Given a query input $x$, $M$ produces a completion $\hat{y} = M(x)$ in a black-box fashion, without revealing any internal parameters or gradients.
The attacker’s goal is to determine whether a particular pair $(x, y)$ was used in training $M$. 

\paragraph{Adversary’s Knowledge}
To ensure fair comparison with prior work~\cite{yang2024gotcha}, we assume a partially informed adversary who has access to a small portion of the training dataset. Specifically, we follow the standard shadow training assumption used in~\cite{yang2024gotcha}, where the adversary knows approximately 20\% of the samples in $ D_{\text{in}} $. This setting reflects practical scenarios where models are trained on a mix of publicly available and proprietary code, making partial data exposure plausible in real-world deployments.

\paragraph{Adversary’s Capabilities}
The adversary is assumed to have black-box access to the target model $ M $, which means that it can query the model with arbitrary code inputs and observe the corresponding outputs, but cannot access internal model details such as architecture, parameters, or training configuration. In our attack setting, the adversary leverages this black-box access to submit clean and perturbed versions of $ x $ and analyze the resulting completions to infer membership. 

\subsection{Overall Approach}
Unlike prior methods that depend on heuristic rules or require training a surrogate model to approximate $M$~\cite{Yang2022,Hisamoto2020}, our approach directly analyzes the victim model’s responses to perturbed queries, thereby simplifying the overall workflow while maintaining strong attack performance. Specifically, we apply adversarial modifications to the input $x$ and examine the variations in $M$’s predictions to reveal membership signals.

\begin{figure*}[htbp]
	\centering
	\includegraphics[width=1.0\textwidth]{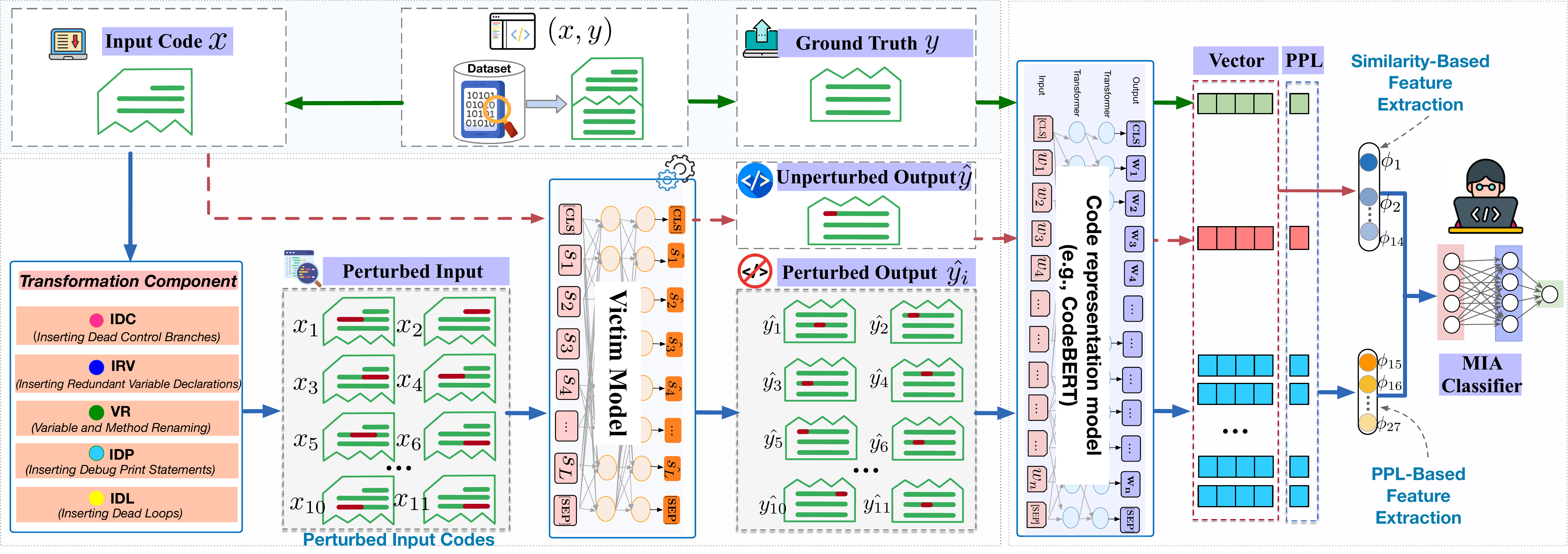}
    \caption{Workflow of the proposed MIA framework, AdvPrompt-MIA 
    }
	\label{fig:MIA_Framework}
\end{figure*}

The workflow of our method is illustrated in Fig.~\ref{fig:MIA_Framework} and consists of the following steps. First, the victim model $M$ is queried with the original input $x$ to obtain the corresponding prediction $\hat{y}$. Next, a set of perturbed inputs $x_i^{\prime}$ is generated by applying small, semantics-preserving modifications to $x$, such as inserting dead code or renaming variables, ensuring that the program’s intended functionality remains unchanged. Each perturbed input $x_i^{\prime}$ is then used to query $M$, resulting in a set of predictions ${\hat{y}_i}$. Subsequently, a pre-trained code embedding model (i.e., CodeBERT~\cite{Feng2020}) is employed to transform $y$, $\hat{y}$, and each $\hat{y}_i$ into vector representations. From these embeddings, similarity- and perplexity-based features are extracted to quantify how closely and consistently $M$’s completions align with the true suffix $y$.
Finally, a binary classifier $G$ is trained on the extracted features using labeled member and non-member samples. During inference, $G$ predicts the membership status of new queries based on their extracted features.

\section{Details of the Proposed Method}
\label{sec:method_detail}
\subsection{Adversarial Perturbations for Membership Exposure}
Adversarial perturbations are key to revealing the memorization behavior of LLMs. As discussed in Section~\ref{sec:our_intuition}, if $(x,y)$ is a training sample, the model's predictions remain stable under small perturbations to $x$; otherwise, the outputs vary more due to weaker learned associations.
This behavioral divergence motivates the construction of \emph{adversarial perturbations} that maintain program semantics while amplifying the prediction stability gap between member and non-member samples, thereby exposing memorization signals that can be exploited by a downstream classifier.

To this end, we design the following five semantics-preserving perturbations. Unlike prior adversarial attack methods that focus on imperceptible changes, our goal is to observe model behavior under meaningful input variations, regardless of whether the modifications are minimal.

\subsubsection{Inserting Dead Control Branches (IDC)}
\label{subsub:irc}

\begin{figure}[htbp]
	\centering
	\includegraphics[width=0.47\textwidth]{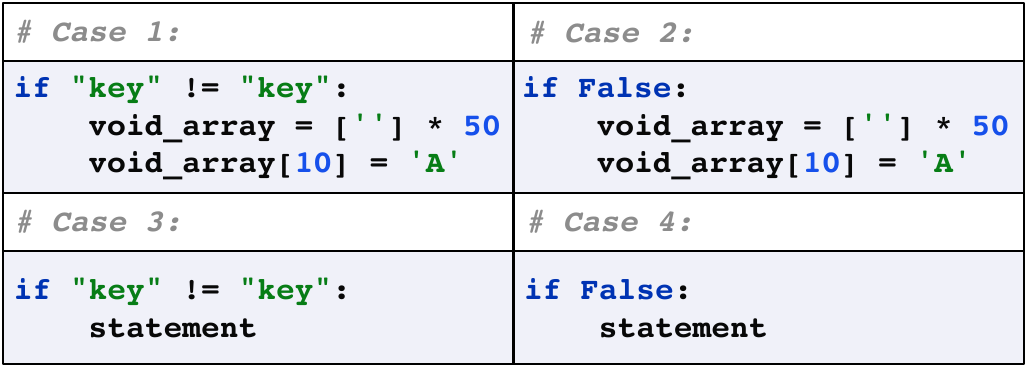}
    \caption{Four strategies for constructing IDC branches
    }
	\label{fig:IDC_forms}
\end{figure}

This transformation inserts a conditional branch with a statically false predicate at a randomly selected line in the code. The inserted branch optionally wraps either an assignment to a non-existent variable or an existing statement from~$x$. Although this modifies the control flow and dependency structure, the runtime behavior remains unaffected since the branch condition is always false and thus never executed. We define four construction strategies, as illustrated in Fig.~\ref{fig:IDC_forms}, and randomly apply two per input during our experiments, as applying all four did not lead to further performance improvements.

Taking Case 3 as an example, the IDC transformation is formally defined as follows.
\begin{equation}
  T_{\text{IDC}} =
  \Bigl\{\,t_{\text{IDC}} \;\bigm|\; 
    \forall x\!\in\!\mathcal{X},\;
    \forall s\!\in\!S(x),\;
    t_{\text{IDC}}(x,s)\!\in\!\mathcal{X}'\Bigr\}
  \label{eq:IDC}
\end{equation}

where $S(x)$ denotes the set of statements in~$x$.

\subsubsection{Inserting Redundant Variable Declarations (IRV)}
\label{subsub:irv}
\begin{figure}[htbp]
	\centering
	\includegraphics[width=0.47\textwidth]{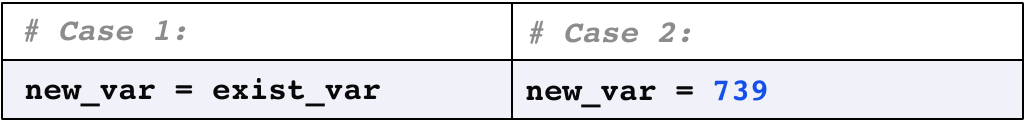}
    \caption{Two forms of the IRV transformation
    }
	\label{fig:IRV_forms}
\end{figure}

This transformation injects a variable declaration that is never used in the subsequent code. The declared variable may be optionally initialized using either an existing variable (Case~1) or a randomly selected constant (Case~2), as shown in Fig.~\ref{fig:IRV_forms}. The insertion is performed after the line where the referenced variable appears; if a constant is used, the insertion position is selected randomly. The formal definition of the IRV transformation for Case~1 is provided in Equation~\ref{eq:IRV}, where $V(x)$ denotes the set of variables in~$x$.

\begin{equation}
  T_{\text{IRV}} =
  \Bigl\{\,t_{\text{IRV}} \;\bigm|\;
    \forall x\!\in\!\mathcal{X},\;
    \forall v\!\in\!V(x):\;
    t_{\text{IRV}}(x,v)\!\in\!\mathcal{X}'\Bigr\},
  \label{eq:IRV}
\end{equation}

\subsubsection{Variable and Method Renaming (VR)}
\label{subsub:vr}

This transformation replaces all occurrences of a variable or method name with a new syntactically valid identifier, typically generated by appending a randomized numeric suffix to the original name. 
The formal definition of the VR transformation is as follows:

\begin{equation}
  T_{\text{VR}} =
  \Bigl\{\,t_{\text{VR}} \;\bigm|\;
    \forall x\!\in\!\mathcal{X},\;
    \forall v\!\in\!V(x):\;
    t_{\text{VR}}(x,v)\!\in\!\mathcal{X}'\Bigr\}.
  \label{eq:VR}
\end{equation}

where $V(x)$  denotes the set of variable and method identifiers in $x$, and $v \in V(x)$ is a selected identifier.
This renaming operation perturbs the token-level representation of the code while preserving its semantics, which has been shown to be effective in prior adversarial attack studies~\cite{zhang2022towards}.

\subsubsection{Inserting Debug Print Statements (IDP)}
\label{subsub:idp}

This transformation, with two cases as shown in Fig.~\ref{fig:IDP_forms}, inserts print statements for logging or debugging purposes, thereby increasing lexical diversity and introducing benign data-flow dependencies. In Case~1, the print statement is inserted at the beginning of the method body; in Case~2, it is placed after a variable declaration. The formal definition of the IDP transformation, corresponding to Case~2, is as follows:

\begin{figure}[htbp]
	\centering
	\includegraphics[width=0.47\textwidth]{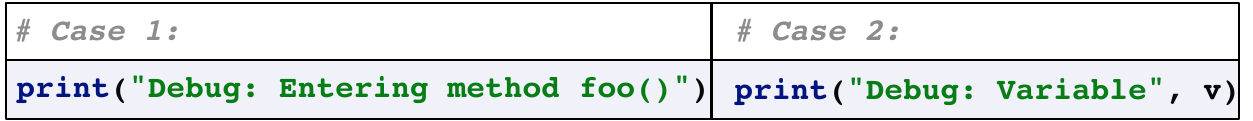}
    \caption{Two forms of the IDP transformation
    }
	\label{fig:IDP_forms}
\end{figure}

\begin{equation}
  T_{\text{IDP}} =
  \Bigl\{\,t_{\text{IDP}} \;\bigm|\;
    \forall x\!\in\!\mathcal{X},\;
    \forall v\!\in\!V(x):\;
    t_{\text{IDP}}(x,v)\!\in\!\mathcal{X}'\Bigr\},
  \label{eq:IRO}
\end{equation}
where $v$ is a variable from $x$ used to establish a data-flow dependency through the inserted print statement.

\subsubsection{Inserting Dead Loops (IDL)}
\label{subsub:idl}

\begin{figure}[htbp]
	\centering
	\includegraphics[width=0.47\textwidth]{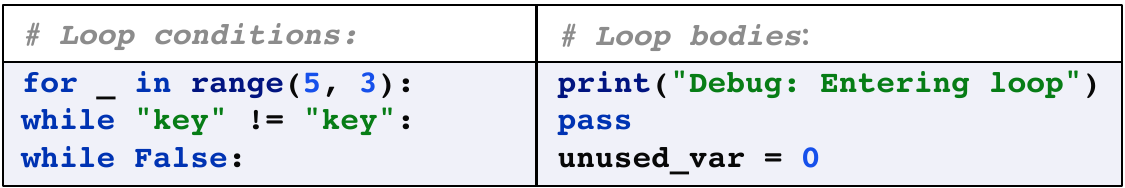}
    \caption{Loop conditions and bodies used in the IDL transformation
    }
	\label{fig:idl_forms}
\end{figure}

This transformation introduces three types of loop conditions that are statically false (i.e., the loops will never execute), optionally populated with a no-op body such as a \texttt{print statement}, \texttt{pass}, or an \texttt{unused assignment}, as illustrated in Fig.~\ref{fig:idl_forms}. The loop is inserted at a randomly selected line in the code to preserve semantics while increasing structural complexity.

The IDL transformation is formally defined as follows:

\begin{equation}
  T_{\text{IDL}} =
  \Bigl\{\,t_{\text{IDL}} \;\bigm|\;
    \forall x\!\in\!\mathcal{X},\;
    t_{\text{IDL}}(x)\!\in\!\mathcal{X}'\Bigr\}.
  \label{eq:IDL}
\end{equation}



Among the above five perturbation strategies, some are inspired by prior adversarial attack works (e.g., VR in ALERT~\cite{yang2022natural} and IDP in DIP~\cite{na2023dip}).
Based on the five perturbation methods, we construct a set of 11 semantically equivalent variants $\mathcal{X}'$ for each input $x$ using Algorithm 1. As shown in the algorithm, each variant is perturbed with one single transformation.
In particular, the resulting set $\mathcal{X}'$ includes two variants each from IDC, IRV, VR, and IDP, and three from IDL. 
All variants preserve the original program semantics, as they do not alter the program’s functional behavior. This is confirmed by our manual verification of representative examples from each transformation.

\begin{algorithm}[ht]
\small
\SetAlgoLined
\SetKwInOut{Input}{Input}
\SetKwInOut{Output}{Output}
\Input{code $x$ with lines $L$, variables $V$, methods $M$}
\Output{perturbed code $x'$}
copy $x_{\mathrm{orig}}\!\leftarrow\!x$; let $\mathcal{X}'\!\leftarrow\!\emptyset$\\
\ForEach{type in \{IDC, IRV, VR, IDP, IDL\}}{
  \uIf{type = IDC}{
    pick $f_1,f_2\in\mathrm{IDC\_forms}$; for each $f\in\{f_1,f_2\}$, apply $t_{\mathrm{IDC}}^{f}$ to $x_{\mathrm{orig}}$, appending result to $\mathcal{X}'$\\
  }
  \uElseIf{type = IRV}{
    for each $g\in\mathrm{IRV\_forms}$, apply $t_{\mathrm{IRV}}^{g}$ to $x_{\mathrm{orig}}$, appending result to $\mathcal{X}'$\\
  }
  \uElseIf{type = VR}{
    pick $h_1,h_2\in\mathrm{VR\_forms}$; for each $h\in\{h_1,h_2\}$, apply $t_{\mathrm{VR}}^{h}$ to $x_{\mathrm{orig}}$, appending result to $\mathcal{X}'$\\
  }
  \uElseIf{type = IDP}{
    pick $p_1,p_2\in\mathrm{IDP\_forms}$; for each $p\in\{p_1,p_2\}$, apply $t_{\mathrm{IDP}}^{p}$ to $x_{\mathrm{orig}}$, appending result to $\mathcal{X}'$\\
  }
  \uElse{
    for each $q\in\mathrm{IDL\_forms}$, apply $t_{\mathrm{IDL}}^{q}$ to $x_{\mathrm{orig}}$, appending result to $\mathcal{X}'$\\
  }
}
\Return{$\mathcal{X}'$}
\caption{Generate the 11 perturbations of $x$.}
\end{algorithm}

\subsection{Training MIA Classifiers}
\label{subsec:training_mia}

\subsubsection{Feature Extraction}
For each original input $x$ and its perturbed variants $\{x_i^{\prime}\}_{i=1}^{11}$, we obtain the model outputs $\hat{y}$ and $\{\hat{y}_i\}_{i=1}^{11}$. We then compute two types of deviations from the reference output $y$: (i) similarity scores and (ii) perplexity values. These measures capture how the model’s predictions vary in response to perturbations and form the basis of the feature vector used for membership inference.

\paragraph{Textual Similarity}
We tokenize and embed each code snippet using CodeBERT~\cite{Feng2020} to obtain a token sequence of length $L$ and an embedding matrix of size $L\times 768$. Applying average pooling yields a single $768$-dimensional vector. Let $v(\cdot)$ denote this embedding process; for any two snippets $y$ and $\hat{y}$, their cosine similarity is defined as:
\begin{equation}
  \mathrm{sim}(y, \hat{y}) \;=\; \frac{v(y)\cdot v(\hat{y})}{\|v(y)\|\|v(\hat{y})\|}.
\end{equation}

\paragraph{Perplexity}
To quantify how anomalous a model’s output is under perturbation, we employ perplexity. For a token sequence $w_1^N = (w_1,\dots,w_N)$, the model assigns conditional probabilities $P(w_i\mid w_{<i})$, and the perplexity of the sequence is defined as:
\begin{equation}
  \mathrm{PPL}(w_1^N)
  \;=\;
  \exp\!\Bigl(-\tfrac{1}{N}\sum_{i=1}^N \ln P(w_i\mid w_{<i})\Bigr)
  \label{eq:perplexity}
\end{equation}
where lower $\mathrm{PPL}$ indicates higher confidence. To evaluate the relative change in perplexity under perturbations, we normalize each perplexity score with respect to the unperturbed output:

\begin{equation}
  \mathrm{PPL}'(\hat{y}_i)
  \;=\;
  \frac{\mathrm{PPL}(\hat{y}_i)\;-\;\mathrm{PPL}(\hat{y})}
       {\mathrm{PPL}(\hat{y})}
  \label{eq:std_perplexity}
\end{equation}

\paragraph{Feature Vector Construction}
For each sample $(x, y)$, we construct a 27-dimensional feature vector that captures both similarity- and perplexity-based cues relevant to membership. Specifically, the vector includes the similarity between the ground-truth suffix $y$ and the model’s unperturbed output $\hat{y}$, followed by the similarities between $y$ and the 11 adversarially perturbed completions $\hat{y}_i$.
To summarize these 12 similarity values, we include their mean $\mu(\mathrm{sim})$ and standard deviation $\sigma(\mathrm{sim})$. In addition, the vector incorporates the standardized perplexity scores $\mathrm{PPL}'(\hat{y}_i)$ for each of the 11 perturbed outputs, along with their mean $\mu(\mathrm{PPL}')$ and standard deviation $\sigma(\mathrm{PPL}')$. Formally, the resulting feature vector is defined as:

\[
\phi(x,y) =
\left[
\begin{aligned}
  &\mathrm{sim}(y,\hat y), \\
  &\;\mathrm{sim}(y,\hat y_1),\;\dots,\;\mathrm{sim}(y,\hat y_{11}),\;\mu(\mathrm{sim}),\;\sigma(\mathrm{sim}),\\
  &\mathrm{PPL}'(\hat y_1),\;\dots,\;\mathrm{PPL}'(\hat y_{11}),\;\mu(\mathrm{PPL}'),\;\sigma(\mathrm{PPL}')
\end{aligned}
\right].
\]
The feature vector captures behavioral differences, serving as the central mechanism that drives our method’s performance.


\subsubsection{Model Training and Inference}
To perform membership inference, we train a multilayer perceptron (MLP) classifier using the high-dimensional feature representations $\phi(x,y)$ constructed in the previous step. Let $D=27$ denote the input layer dimension, which matches the dimensionality of the feature vector. The network comprises three hidden layers of size 512 and an output layer that produces logits in $\mathbb{R}^C$, where $C=2$ for binary membership classification.
Each hidden layer performs a linear mapping with ReLU activation:

\begin{equation}
  \mathbf{h}_i = \mathrm{ReLU}\bigl(\mathbf{W}_i\,\mathbf{h}_{i-1} + \mathbf{b}_i\bigr),\quad
  \mathbf{h}_0 = \phi(x,y),
\end{equation}
followed by a dropout operation to mitigate overfitting. After the final hidden layer $\mathbf{h}_3$, the output logits are computed as
\begin{equation}
  \mathbf{z} = \mathbf{W}_{\mathrm{out}}\,\mathbf{h}_3 + \mathbf{b}_{\mathrm{out}},
  \quad
  \mathbf{z}\in\mathbb{R}^C.
\end{equation}
The classifier is trained using standard cross-entropy loss.

At inference time, given a data pair $(x, y)$, we first obtain the unperturbed output $\hat{y} = M(x)$ from the victim model. We then apply semantics-preserving perturbations to $x$ to generate a set of modified inputs $\{x_i^{\prime}\}$ and collect their corresponding outputs $\{\hat{y}_i\}$. Based on these outputs and the reference $y$, we compute the feature vector $\phi(x, y)$ as described in Section~\ref{subsec:training_mia}. Finally, this feature vector is fed into the trained MLP classifier to determine the membership status.
This end-to-end pipeline relies solely on the victim model’s outputs under controlled perturbations and the feature vectors designed to capture output variations indicative of memorization.

\section{Experiment Design}
\label{eval_setting}
We aim to answer the following research questions (RQs)\footnote{Code is available at \url{https://github.com/YuanJiangGit/MIA_Adv}}:

RQ1: To what extent does the proposed MIA method outperform state-of-the-art baselines?

RQ2: Which perturbation strategies and feature components contribute most to attack performance?

RQ3: To what extent can the proposed method generalize to other code LLMs beyond Code Llama 7B?

RQ4: Can the proposed method train an MIA classifier on one model that transfers effectively to other target models?

RQ5: To what extent does the proposed method generalize with varying or no knowledge of target training datasets?

\subsection{Dataset}
\label{sec:dataset}

\vspace{0.3em}
Code generation (or code completion) is a core capability of code LLMs and provides a representative setting for studying membership inference~\cite{yang2024gotcha}. We adopt APPS and HumanEval, two widely used benchmarks, as evaluation datasets for assessing MIA performance. To investigate potential leakage between our evaluation benchmarks and the pre-training corpora of the target models, we perform a text-based matching analysis. 
Among the five victim models considered in this paper (introduced in Section~\ref{sec:eval_setting}), StarCoder2 is the only one with publicly traceable pre-training data
(The Stack v2, 32.1TB, 1.15B files)~\cite{lozhkov2024starcoder}.
We therefore conduct matching experiments against this corpus and find no exact overlaps with the code samples from HumanEval and APPS, reducing the likelihood of data leakage.
Moreover, both benchmarks are widely adopted as official test sets for evaluating the functional correctness of target models~\cite{roziere2023code,lozhkov2024starcoder,guo2024deepseek}, further supporting their exclusion from pre-training data to ensure fair evaluation.

APPS~\cite{hendrycks2021measuring} contains 5\,000 training tasks and 5\,000 testing tasks.  
For each task, we extract a single reference implementation and discard any tasks without valid solutions. This yields 5\,000 training examples for constructing the victim model’s training dataset $ D_{\mathrm{in}} $, and 3{,}765 testing examples, which serve as the basis for the non-member dataset $ D_{\mathrm{out}} $.
HumanEval~\cite{chen2021evaluating} consists of 164 hand-crafted programming problems.  
Following a similar setup, we randomly divide the problems into two disjoint subsets of equal size. One subset (82 problems) is used to construct $ D_{\mathrm{in}} $, while the remaining 82 problems form $ D_{\mathrm{out}} $. Each problem is associated with a single reference solution. Note that $D_{\mathrm{in}}$ is used to fine-tune publicly released code LLMs rather than train from scratch.

We adopt a \emph{partial knowledge} threat model, in which the adversary has access to only a small portion of the training data.  
Specifically, 20\% of $ D_{\mathrm{in}} $ is assumed to be known to the adversary and is used as the positive class when training the MIA classifier.  
An equal number of non-member samples is randomly drawn from $ D_{\mathrm{out}} $ to form a balanced training set. This setting, including the exact ratio (i.e., 20\%), is consistent with prior work~\cite{yang2024gotcha}.
For evaluation, the same sampling strategy is applied to the remaining 80\% of $ D_{\mathrm{in}} $ and to the rest of $ D_{\mathrm{out}} $, ensuring that training and evaluation are performed on disjoint data splits derived from the original benchmarks.

\begin{figure}[htbp]
	\centering
	\includegraphics[width=0.5\textwidth]{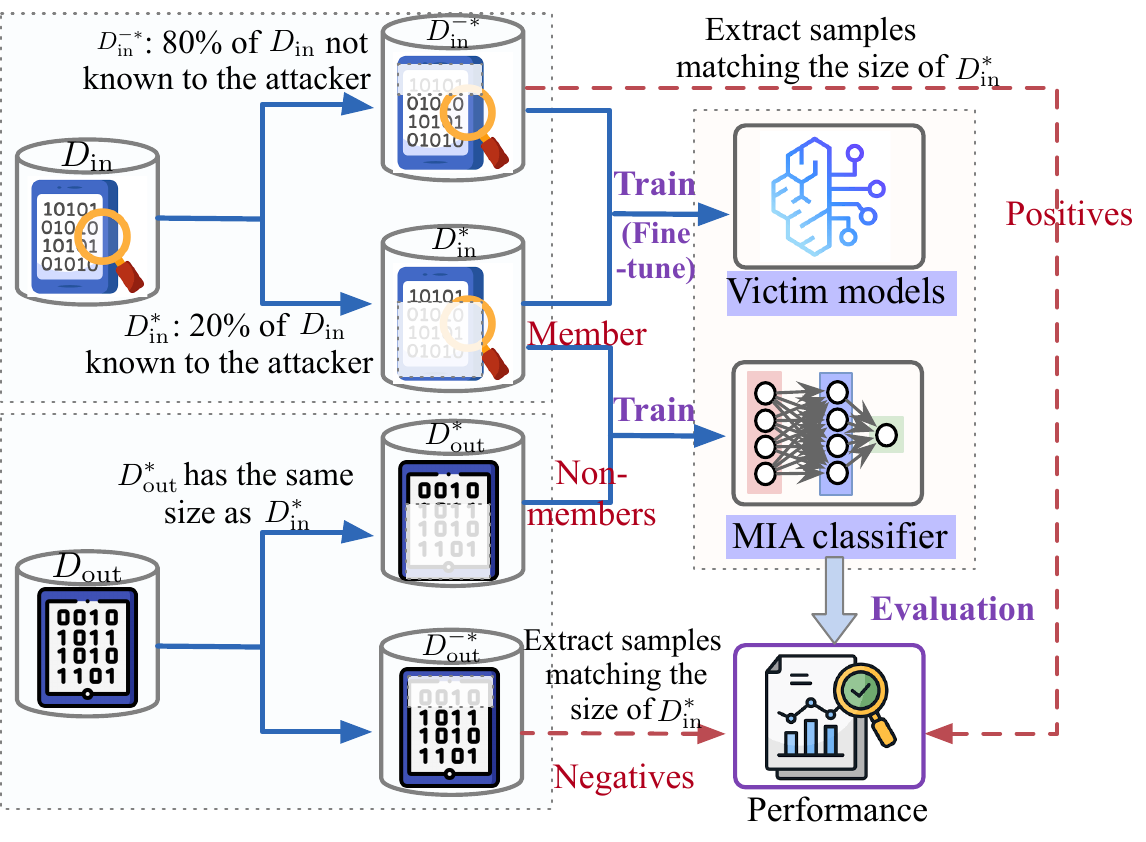}
    \caption{Workflow for dataset construction used in experiments
    }
	\label{fig:data_pipeline}
\end{figure}

The data preparation process is illustrated in Fig.~\ref{fig:data_pipeline}, where $ D_{\mathrm{in}} $ and $ D_{\mathrm{out}} $ denote the training and testing sets, used to simulate member and non-member samples, respectively.

\subsection{Evaluation Metrics}
Consistent with prior work~\cite{Hisamoto2020,yang2024gotcha}, we adopt three metrics to evaluate the performance of the MIA classifier: True Positive Rate (TPR), False Positive Rate (FPR), and Area Under the ROC Curve (AUC). These are defined as: 
TPR = $\mathrm{TP} / (\mathrm{TP} + \mathrm{FN})$, 
FPR = $\mathrm{FP} / (\mathrm{FP} + \mathrm{TN})$, and 
AUC = $\int_0^1 \mathrm{TPR}(\tau) \, \mathrm{dFPR}(\tau)$,
where $\mathrm{TP}$, $\mathrm{FP}$, $\mathrm{TN}$, and $\mathrm{FN}$ represent true positives, false positives, true negatives, and false negatives, respectively, and $\tau$ denotes the decision threshold.

We choose these metrics because TPR reflects the attacker’s ability to correctly identify member samples (i.e., attack power), FPR captures the rate of incorrect membership predictions (i.e., attack error), and AUC provides a threshold-independent summary of the trade-off between them, offering an overall assessment of classifier performance.





%

\subsection{Baseline Methods}
\label{sec:baseline}
We compare our method against four representative black- or gray-box MIA baselines recently proposed for evaluating membership inference risks in target code LLMs, where GOTCHA and GT-Match have been detailed in Section~\ref{sec:MIA_define}.

GOTCHA~\cite{yang2024gotcha} trains a shadow model on a small known portion of the training data, uses its outputs to train a binary classifier, and transfers it to the victim model for inference.

GT-Match~\cite{al2024traces} labels a sample as a member if the model generates a suffix that exactly matches the ground truth, leveraging verbatim memorization.

PPL-Rank~\cite{Yang2024} ranks samples by token-level perplexity and predicts the top 50\% as members, assuming lower perplexity for training data.

BUZZER-b~\cite{carlini2021extracting} applies lowercase perturbations and measures the change in representation, detecting memorization based on casing sensitivity.

\subsection{Evaluation Setting}
\label{sec:eval_setting}
We evaluate our proposed MIA method on the widely used Code Llama 7B model~\cite{roziere2023code}. To examine its generalizability, we further conduct experiments on additional code LLMs, including Deepseek-Coder 7B~\cite{guo2024deepseek}, StarCoder2 7B~\cite{lozhkov2024starcoder,li2023starcoder}, Phi-2 2.7B~\cite{javaheripi2023phi}, and WizardCoder 7B~\cite{luo2023wizardcoder}.
Following prior work and common practices~\cite{he2024instruction}, we fine-tune all code LLMs for 5 epochs using a learning rate of 2e-5, Adam optimizer (weight decay 1e-2, $\epsilon$=1e-8), batch size of 2, and 4-step gradient accumulation. Phi-2 (2.7B) is fully fine-tuned, while all 7B models are fine-tuned using LoRA ($r=16$, $\alpha$=32, dropout=0.1). We also perform a random search over alternative settings and find that this configuration consistently yields the best overall functional correctness across all models.

Our MIA classifier is implemented as a fully connected three-layer neural network with a hidden size of 512. It is optimized using the Adam optimizer~\cite{kingma2014adam}, with a learning rate of 1e-3 and no weight decay, and trained for 25 epochs using randomly shuffled mini-batches of size 4.
The model architecture and hyperparameters are selected via random search on the HumanEval dataset. All experiments are conducted on two NVIDIA A6000 GPUs, each equipped with 48 GB of memory.

For baseline comparisons, we use the original implementations where available (e.g., GOTCHA and BUZZER-b), and re-implement the methods when necessary. For rank-based approaches such as PPL-Rank and BUZZER-b, we classify the top 50\% of the samples as members, consistent with the balanced member/non-member split in the evaluation datasets and the setting used in prior work~\cite{yang2024gotcha}.


\section{Experiment Results}
\label{eval_results}
\subsection{RQ1: To what extent does the proposed MIA method outperform state-of-the-art baselines?}
\label{sec:rq1}
To assess the effectiveness of our proposed MIA method, we compare it against four baselines using the Code Llama 7B model. All experiments are conducted on the APPS and HumanEval datasets (see Section~\ref{sec:dataset} for dataset details), and the baselines are described in Section~\ref{sec:baseline}. 
For each method, we report TPR, FPR, and AUC.

Table~\ref{tab:rq1-results} presents the performance comparison between our method, AdvPrompt-MIA, and existing baselines. As shown, our method consistently outperforms all baselines across both datasets, demonstrating its effectiveness regardless of dataset scale. Specifically, on the relatively large-scale APPS dataset, AdvPrompt-MIA achieves AUC improvements of 67.2\%, 86.3\%, 63.8\% and 90.0\% over GOTCHA, GT-Match, PPL-Rank and BUZZER-b, respectively. The performance gap is even more pronounced on the smaller HumanEval dataset, where our method achieves 69.6\%, 73.2\%, 102.1\% and 94.0\% higher AUC than the same baselines.

\begin{table}[htbp]
  \centering
  \caption{Comparison of the Proposed Method and Baselines for Membership Inference Attacks on Code Llama 7B}
  \label{tab:rq1-results}
  \setlength{\tabcolsep}{8pt}
  \renewcommand{\arraystretch}{1.1}
  \resizebox{0.47\textwidth}{!}{
  \begin{tabular}{@{}llccc@{}}
    \toprule
    \textbf{Dataset}      & \textbf{Method}     & \textbf{TPR}\,$\uparrow$ & \textbf{FPR}\,$\downarrow$ & \textbf{AUC}\,$\uparrow$ \\
    \midrule
    \multirow{5}{*}{HumanEval}
      & GOTCHA           & 0.35 & 0.40 & 0.58 \\
      & GT-Match         & 0.55 & 0.40 & 0.56 \\
      & PPL-Rank         & 0.40 & 0.55 & 0.48 \\
      & BUZZER-b         & 0.48 & 0.50 & 0.50 \\
      & AdvPrompt-MIA    & \textbf{0.85} & \textbf{0.05} & \textbf{0.97} \\
    \midrule
    \multirow{5}{*}{APPS}
      & GOTCHA           & 0.36 & 0.40 & 0.56 \\
      & GT-Match         & 0.56 & 0.58 & 0.51 \\
      & PPL-Rank         & 0.58 & 0.42 & 0.58 \\
      & BUZZER-b         & 0.52 & 0.50 & 0.50 \\
      & AdvPrompt-MIA    & \textbf{0.90} & \textbf{0.14} & \textbf{0.95} \\
    \bottomrule
  \end{tabular}
  }
\end{table}

Our method achieves strong performance primarily by leveraging a series of semantics-preserving perturbations to amplify membership signals. It then constructs feature representations that capture the consistency or variability in the model’s responses to these perturbations. These features enable the deep learning classifier to distinguish member from non-member instances by identifying robust behavioral patterns, even in the absence of exact output matches.

In addition, we observe that all four baseline methods exhibit limited effectiveness on both datasets. 
GOTCHA suffers from severe overfitting because only 20\% of the victim’s training data is used as member samples; the resulting shadow model fails to mimic the victim model’s behavior, which prevents the MIA classifier from learning meaningful features and leads to a high FPR and limited effectiveness.
GT-Match exhibits similar performance to GOTCHA and remains ineffective. Although this method has been widely used to detect the leakage of sensitive information such as API keys or credentials~\cite{alkaswan2024traces}, it struggles to accurately infer the membership status of functional code, which is the primary content of the APPS and HumanEval datasets. PPL-Rank also fails to distinguish members from non-members, likely because perplexity alone cannot reliably indicate memorization; in fact, recent work shows that LLMs can generate code with higher quality than human-written samples~\cite{jiang2025enhancing}. BUZZER-b, while introducing input perturbations to detect memorization, applies only a single perturbation, i.e., converting tokens to uppercase, which may be insufficient to trigger observable differences, especially in models robust to such surface input changes.

\conclbox{
\textbf{Conclusion}: AdvPrompt-MIA outperforms state-of-the-art baselines on both datasets, demonstrating that semantics-preserving perturbations, together with features derived from the resulting outputs, effectively amplify memorization signals and enable more accurate MIA.
}

\subsection{RQ2: Which perturbation strategies and feature components contribute most to attack performance?}
\label{sec:rq2}
\begin{figure}[htbp]
	\centering
	\includegraphics[width=0.49\textwidth]{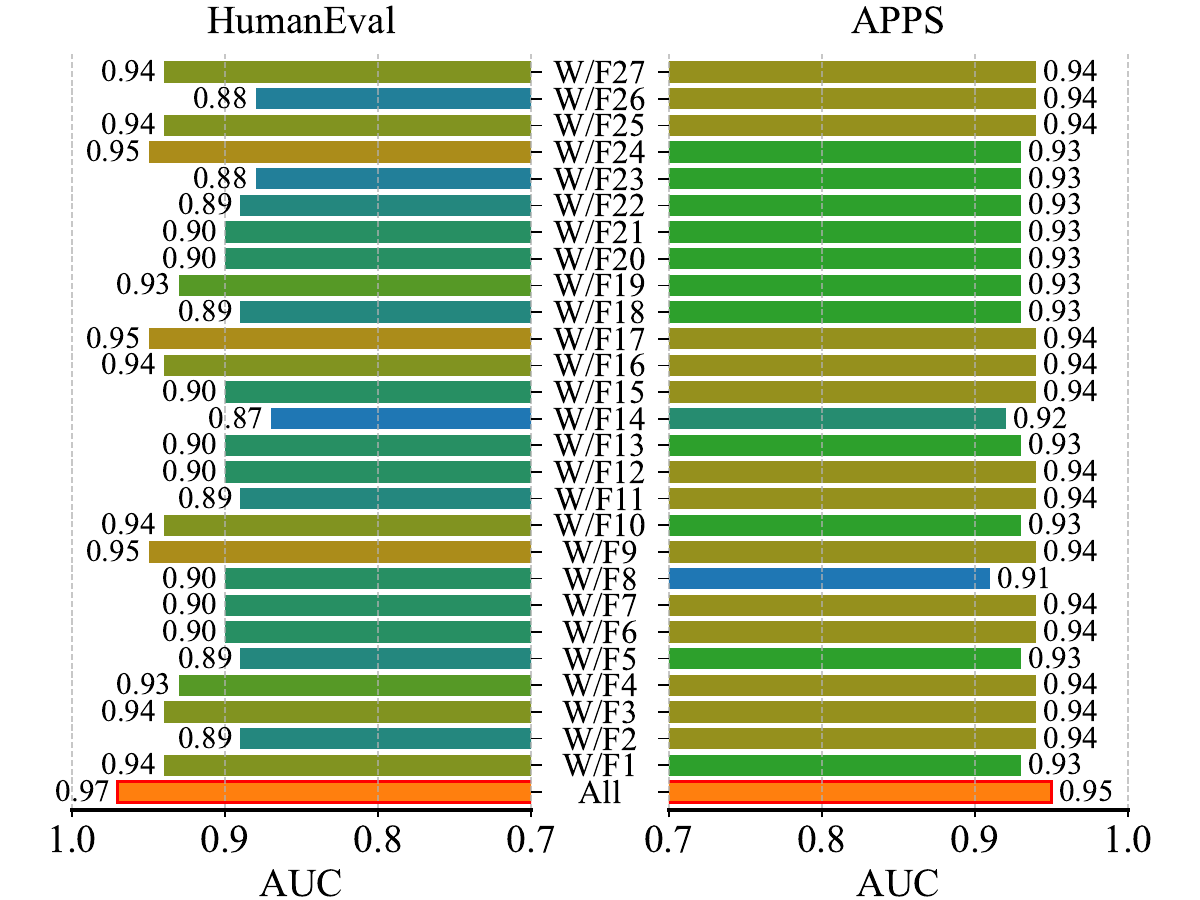}
	\caption{AUC between our method (using all features) and its variants with each of the 27 features removed,
where W/F{i} denotes that the i-\textit{th} feature is excluded when constructing the feature vectors for the MIA classifier.}
	\label{fig:ablation_features}
\end{figure}

\begin{figure}[htbp]
	\centering
	\includegraphics[width=0.45\textwidth]{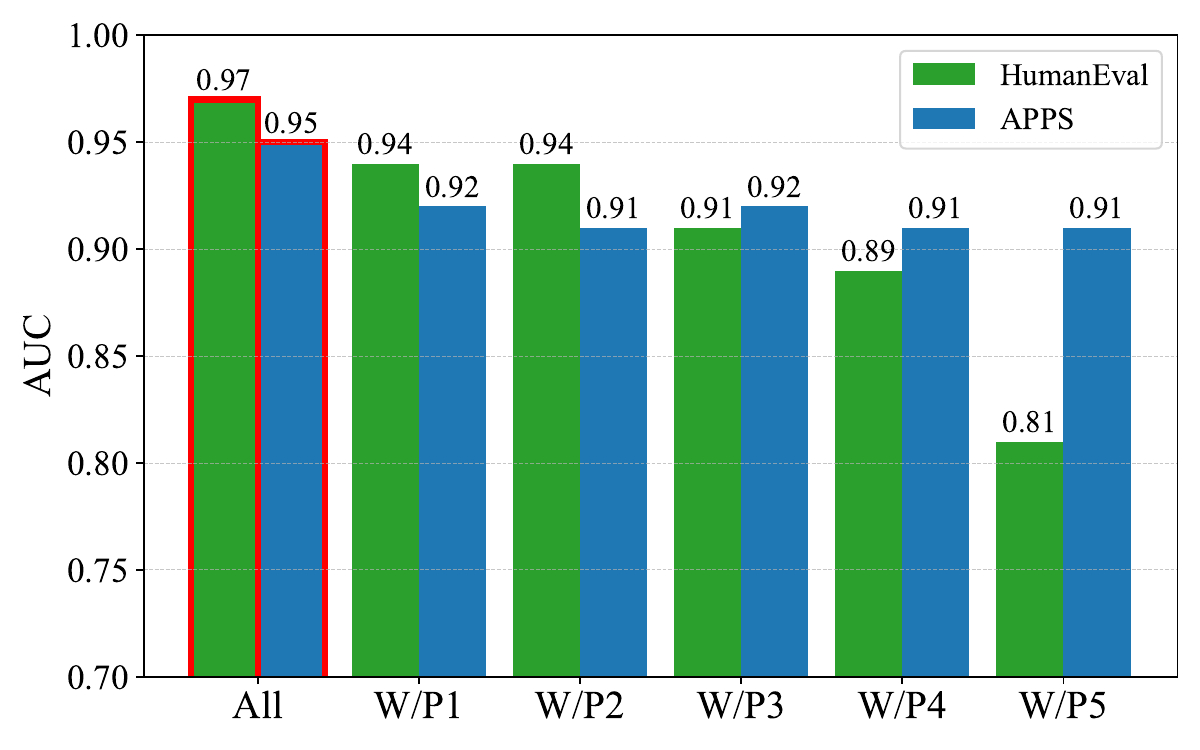}
	\caption{AUC between our method (i.e., All) and its variants with each of the five perturbations removed,
where W/P{i} denotes that the i-\textit{th} perturbation is excluded during adversarial prompt generation.}
	\label{fig:ablation_perturb}
\end{figure}

To assess the contribution of each component in our approach, we perform an ablation study on both the five semantics-preserving perturbation strategies and the 27-dimensional feature vector introduced in Section~\ref{sec:method_detail}. Specifically, we evaluate two settings: (i) removing one feature dimension at a time, and (ii) disabling one perturbation strategy at a time, while retraining the MIA classifier for each variant. All other experimental configurations are kept consistent with those used in Section~\ref{sec:rq1}. The effect of omitting individual feature dimensions is illustrated in Fig.~\ref{fig:ablation_features}, and the performance impact of excluding each perturbation is shown in Fig.~\ref{fig:ablation_perturb}.

As shown in Fig.~\ref{fig:ablation_features}, our full method consistently outperforms all ablated variants in terms of AUC, with performance drops ranging from 2.1\% to 10.3\% on HumanEval and from 1.1\% to 4.2\% on APPS when individual features are removed. This result demonstrates the contribution of each feature dimension to the overall effectiveness of the MIA classifier. Notably, the 14th feature, $\sigma(\mathrm{sim})$, is the most impactful, as its removal leads to the largest average performance drop across both datasets. This finding suggests that the variability in similarity scores plays a key role in distinguishing member from non-member samples, as it reflects the degree of output stability. The result is consistent with our hypothesis that member samples yield more stable completions under perturbations, while non-member samples exhibit greater variability.

Fig.~\ref{fig:ablation_perturb} further illustrates the contribution of each perturbation strategy by reporting the AUC when each is individually removed. The results show performance drops ranging from 3.1\% to 16.5\% on HumanEval and from 3.2\% to 4.2\% on APPS.
Among all perturbations, the fifth strategy (i.e., IDL) leads to the most significant degradation when removed. This may be because IDL introduces structurally valid yet semantically inert code blocks, which more effectively expose discrepancies in the model’s generalization behavior between member and non-member inputs. 
These findings underscore the importance of both perturbation diversity and feature expressiveness in enabling reliable membership inference.

\conclbox{
\textbf{Conclusion}: The high performance of our method is attributed to the individual contributions of each feature dimension in the designed vector and the effectiveness of the proposed perturbation strategies.
}

\subsection{RQ3: To what extent can the proposed method generalize to other code LLMs beyond Code Llama 7B?}
In RQ1, we demonstrate that our method achieves superior performance compared to state-of-the-art baselines on the Code Llama 7B model. To further assess its generalizability, we evaluate the proposed approach on four additional widely used code LLMs: Deepseek-Coder 7B~\cite{guo2024deepseek}, StarCoder2 7B~\cite{lozhkov2024starcoder}, Phi-2 2.7B~\cite{javaheripi2023phi}, and WizardCoder 7B~\cite{luo2023wizardcoder}. 
For each victim model, we conduct experiments on both the HumanEval and APPS benchmarks, following the same data preparation and evaluation protocol described in Section~\ref{sec:dataset}. The experimental results are summarized in Table~\ref{tab:rq3-results}.

\begin{table}[htbp]
  \centering
  \caption{Effectiveness of the Proposed Method across Multiple Code LLMs on HumanEval and APPS}
  \label{tab:rq3-results}
  \setlength{\tabcolsep}{8pt}
  \renewcommand{\arraystretch}{1.1}
  \resizebox{0.47\textwidth}{!}{
  \begin{tabular}{@{}llccc@{}}
    \toprule
    \textbf{Dataset}     & \textbf{Victim model}   & \textbf{TPR}\,$\uparrow$ & \textbf{FPR}\,$\downarrow$ & \textbf{AUC}\,$\uparrow$ \\
    \midrule
    \multirow{4}{*}{HumanEval}
      & Deepseek-Coder 7B      & 0.80 & 0.10 & 0.91 \\
      & StarCoder2 7B          & 0.75 & 0.10 & 0.92 \\
      & Phi-2 2.7B             & 0.90 & 0.25 & 0.92 \\
      & WizardCoder 7B         & 0.75 & 0.05 & 0.95 \\
    \midrule
    \multirow{4}{*}{APPS}
      & Deepseek-Coder 7B      & 0.82 & 0.19 & 0.89 \\
      & StarCoder2 7B          & 0.84 & 0.16 & 0.91 \\
      & Phi-2 2.7B             & 0.79 & 0.21 & 0.85 \\
      & WizardCoder 7B         & 0.76 & 0.25 & 0.81 \\
    \bottomrule
  \end{tabular}
  }
\end{table}

As shown in Table~\ref{tab:rq3-results}, our method, AdvPrompt-MIA, consistently exhibits strong attack performance across all evaluated code models. For example, on HumanEval, it achieves AUC scores above 0.91, indicating robust generalization.
This robust performance further supports our core assumption: for diverse code LLMs, memorized training samples tend to yield stable model outputs under small, semantics-preserving perturbations, whereas non-member samples result in more significant variations. These behavioral differences provide discriminative signals that enable effective membership inference.

AdvPrompt-MIA is explicitly designed to leverage this assumption by capturing perturbation-induced behavioral cues. Through a combination of semantics-preserving code transformations and a learning-based classification framework, our method automatically identifies internal consistency signals indicative of memorization, without requiring access to the model’s architecture or training details. This design enables AdvPrompt-MIA to generalize well and maintain stable performance across a wide range of code LLMs.

\conclbox{
\textbf{Conclusion}: AdvPrompt-MIA generalizes well across diverse code LLMs, consistently achieving strong performance regardless of the target model’s internal design.
}

\subsection{RQ4: Can the proposed method train an MIA classifier on one model that transfers effectively to other target models?}
\begin{figure}[htbp]
	\centering
	\includegraphics[width=0.5\textwidth]{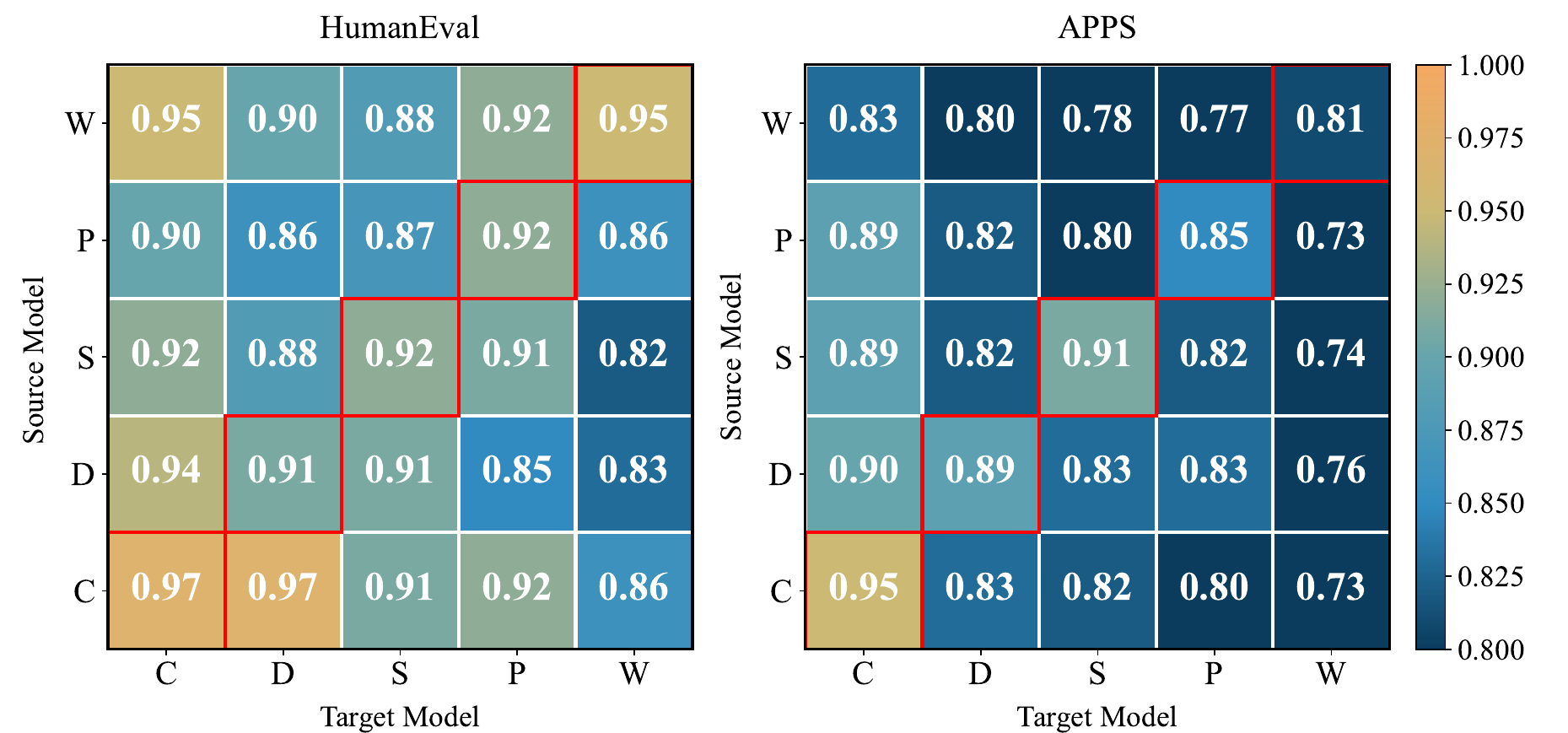}
	\caption{Cross-model AUC scores of MIA classifiers trained on source models and evaluated on target models. C, D, S, P, and W refer to Code Llama, Deepseek-Coder, StarCoder2, Phi-2, and WizardCoder, respectively.
    }
	\label{fig:transfer_models_heatmap}
\end{figure}

To further assess the transferability of our method and examine whether perturbation-induced features generalize across different code LLMs, we conduct cross-model experiments. Specifically, we select pairs of models from Code Llama and those evaluated in RQ3. For each experiment, we use one model to generate training vectors for the MIA classifier using our method, and directly apply the trained classifier to perform membership inference on another model, without any retraining or adaptation. All experiments are conducted on the HumanEval and APPS benchmarks, following the same data preparation process described in Section~\ref{sec:dataset}. For each source-target model pair, we report AUC scores, as shown in Fig.~\ref{fig:transfer_models_heatmap}.

As shown in Fig.~\ref{fig:transfer_models_heatmap}, our method consistently achieves strong performance across all model pairs on both datasets, with the best AUC reaching 0.97 when the MIA classifier is trained on Code Llama and evaluated on Deepseek-Coder on HumanEval and 0.90 when trained on Deepseek-Coder and evaluated on Code Llama on APPS. These results suggest that the perturbation-induced feature space learned from one model transfers well to others, even when the models differ significantly in architecture or training details.

We attribute this strong transferability to the shared behavioral property of code LLMs. Specifically, for member samples, semantics-preserving perturbations result in stable outputs that remain close to the ground-truth completion, whereas for non-members, the same perturbations cause greater variability. Our method captures these behavioral patterns by constructing feature vectors from the outputs of perturbed inputs. The fact that a classifier trained on one model can accurately infer membership on another suggests that these feature vectors, despite originating from different models, occupy a similar space. This similarity arises not from architectural alignment, but from the underlying consistency in how different LLMs respond to membership under perturbation.

The similar behavioral patterns exhibited by different models under perturbations allow AdvPrompt-MIA to generalize well across models. However, subtle differences in how individual models respond to the same perturbations can cause certain models to achieve higher performance. 
For example, the proposed semantics-preserving perturbations tend to induce more regular and stable output behaviors on Code Llama, likely owing to its large-scale and diverse pre-training data, whereas such effects are comparatively weaker on other models. These clearer behavioral patterns, when encoded into our feature vectors, allow the MIA classifier to more easily distinguish whether its input is a member sample, thereby improving inference performance even when the classifier is trained on other models.
This explains Fig.~\ref{fig:transfer_models_heatmap}, where (1) some target models (e.g., Code Llama) consistently outperform others, and (2) in rare cases—1 of 20 on HumanEval and 3 on APPS—the transfer setting yields higher AUC than the corresponding non-transfer baseline (e.g., Deepseek-Coder$\rightarrow$Code Llama vs. Deepseek-Coder$\rightarrow$Deepseek-Coder).

\begin{figure}[htbp]
	\centering
	\includegraphics[width=0.43\textwidth]{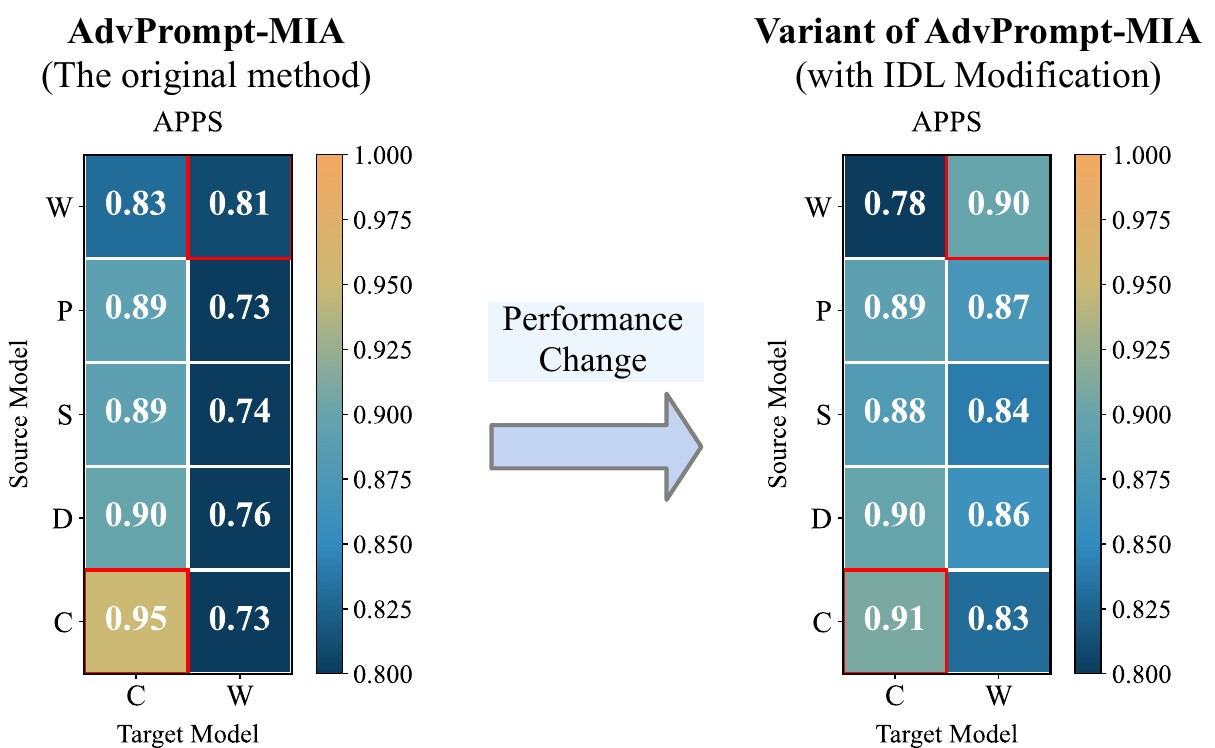}
	\caption{Cross-model performance changes of our method after modifying only the IDL transformation (C, D, S, P, and W refer to Code Llama, Deepseek-Coder, StarCoder2, Phi-2, and WizardCoder, respectively).
    }
	\label{fig:IDL_impact_transfer}
\end{figure}

To verify whether the favorable performance of our method on certain target models stems from the specific perturbations employed, we take the IDL transformation as an example and revise it by replacing the simple loop body with more sophisticated code snippets, provided in the replication package\footnote{\url{https://github.com/YuanJiangGit/MIA_Adv/blob/dev/Modified_IDL.md}} for brevity. The revised IDL remains semantics-preserving, as all conditions are false and loop bodies are never executed. We then compare the performance changes of our method on two target models, Code Llama and WizardCoder, before and after this modification. These two are chosen because they represent the overall best- and worst-performing target models on APPS. As shown in Fig.~\ref{fig:IDL_impact_transfer}, performance decreases on the target model Code Llama but improves significantly on WizardCoder. These findings show that although our method performs well across all victim models, perturbations affect them differently, highlighting adaptive model-specific perturbation optimization as a promising direction for future work.

\conclbox{
\textbf{Conclusion}: AdvPrompt-MIA effectively captures transferable behavioral patterns, enabling a classifier trained on one model to infer membership on unseen models with strong performance. However, its transferability on specific targets is affected by the applied transformations.
}

\subsection{RQ5: To what extent does the proposed method generalize with varying or no knowledge of target training datasets?}

\begin{figure}[htbp]
	\centering
	\includegraphics[width=0.45\textwidth]{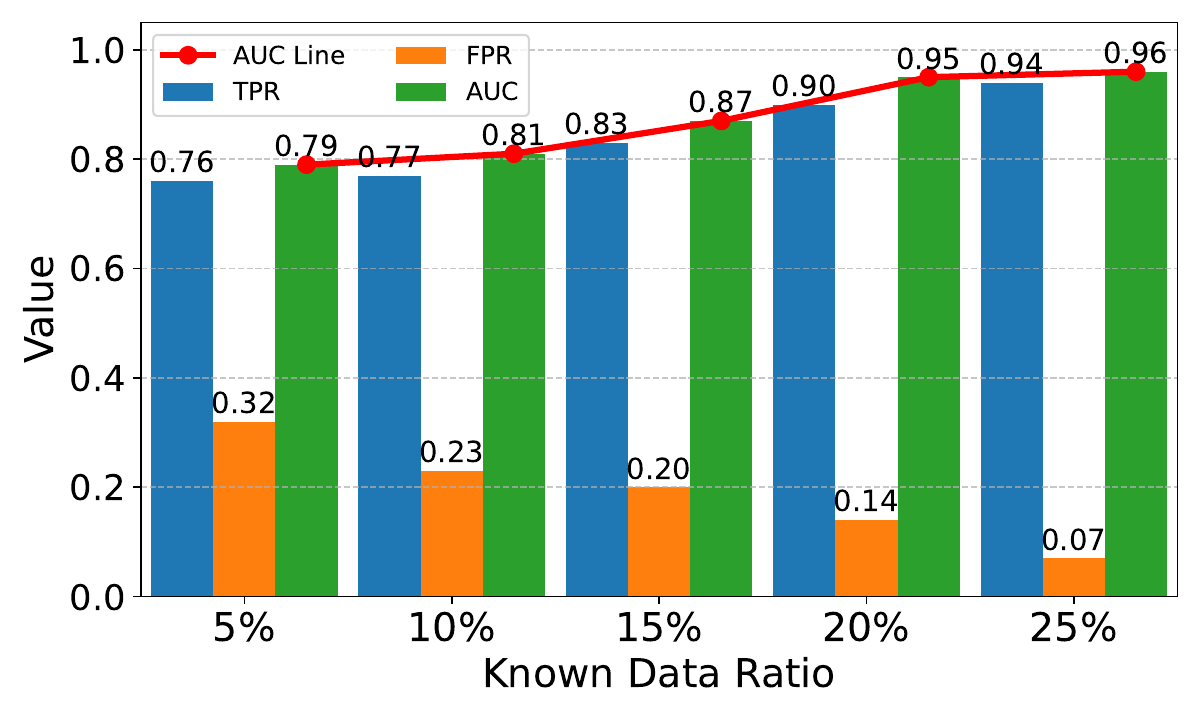}
	\caption{Sensitivity of AdvPrompt-MIA to Known Data Proportions.
    }
	\label{fig:dataset_generalize}
\end{figure}

Our method AdvPrompt-MIA assumes that approximately 20\% of the member samples used to train the victim model are available to the attacker. This assumption is plausible in real-world settings, where code LLMs are often trained on a mixture of public repositories and proprietary code. Moreover, this partial knowledge setup is consistent with prior work~\cite{yang2024gotcha}. To evaluate the sensitivity of our method to the adversary’s knowledge, we vary the ratio of known data from 5\% to 25\% in increments of 5\% and conduct experiments on Code Llama 7B using APPS, which is larger than HumanEval and more suitable for small known-data settings. 
The results in Fig.~\ref{fig:dataset_generalize} show that performance improves by about 0.05 AUC on average for every 5\% increase in known data between 5\% and 20\%, then stabilizes beyond 20\%. Even with only 5\% known data, AdvPrompt-MIA achieves 0.79 AUC, demonstrating its effectiveness under limited adversary knowledge. This robustness arises from the clear behavioral differences exhibited by code LLMs when handling member and non-member samples under the designed perturbations, which can be reliably captured even with a small amount of known data.


To further assess the generalizability of our method, we also consider a more challenging setting in which the attacker has no access to member samples from the target dataset. Specifically, we evaluate whether an MIA classifier trained on one dataset via our method can be transferred to a different dataset, while keeping the target code LLM unchanged. For each victim model, the classifier is trained using member and non-member samples from one dataset (e.g., HumanEval), and evaluated on another (e.g., APPS). The AUC scores across all training–evaluation dataset pairs are summarized in Table~\ref{tab:model_auc}.


\begin{table}[htbp]
\centering
\caption{AUC Performance of MIA Classifier with Training and Evaluation Performed on Different Datasets}
\label{tab:model_auc}
\resizebox{0.47\textwidth}{!}{
\begin{tabular}{lllc}
\toprule
\textbf{Victim models} & \textbf{Training Dataset} &  \textbf{Evaluation Dataset}  & \textbf{AUC}\,$\uparrow$ \\
\midrule
Code Llama 7B            & HumanEval & HumanEval  & 0.97 \\
                            & HumanEval & APPS       & 0.78 \\
                            & APPS      & HumanEval  & 0.94 \\
\midrule
Deepseek-Coder 7B        & HumanEval & HumanEval  & 0.91 \\
                            & HumanEval & APPS       & 0.78 \\
                            & APPS      & HumanEval  & 0.86 \\
\midrule
StarCoder2 7B            & HumanEval & HumanEval  & 0.92 \\
                            & HumanEval & APPS       & 0.74 \\
                            & APPS      & HumanEval  & 0.95 \\
\midrule
Phi-2 2.7B               & HumanEval & HumanEval  & 0.92 \\
                            & HumanEval & APPS       & 0.73 \\
                            & APPS      & HumanEval  & 0.85 \\
\midrule
WizardCoder 7B           & HumanEval & HumanEval  & 0.95 \\
                            & HumanEval & APPS       & 0.68 \\
                            & APPS      & HumanEval  & 0.87 \\
\bottomrule
\end{tabular}}
\end{table}

As shown in Table~\ref{tab:model_auc}, our method remains effective even without access to member samples from the target dataset.
For example, when targeting Code Llama, training the MIA classifier on APPS and evaluating it on HumanEval yields an AUC of 0.94, close to the within-dataset result (i.e., training and evaluating on HumanEval). This demonstrates the strong cross-dataset transferability of AdvPrompt-MIA.

Moreover, we observe an asymmetric trend: when the MIA classifier is trained on APPS and tested on HumanEval, performance is consistently better than the reverse. This may be attributed to the larger size and diversity of the APPS dataset, which provides a richer training signal. These findings suggest that, in practice, when the target dataset is unknown, training the MIA classifier on a larger and more diverse dataset can lead to better cross-dataset inference performance.

\conclbox{
\textbf{Conclusion}: AdvPrompt-MIA demonstrates robustness under limited adversary knowledge and shows strong cross-dataset transferability, which can be further enhanced by training the MIA classifier on larger datasets.
}

\section{Related Work}

\label{sec:related_work}
Existing work~\cite{chen2024promise,rabin2023memorization,carlini2022quantifying,huang2024your,finkman2024codecloak} has investigated the memorization behavior of code LLMs to reveal potential risks of data leakage. 
Although MIAs are well explored in machine learning~\cite{he2025towards,choquette2021label}, they are still emerging for code LLMs. Following~\cite{yang2024survey}, we classify them as gray- or black-box attacks.

\paragraph{Gray-Box MIA}
Gray-box adversaries cannot access the model’s architecture, parameters, or gradients~\cite{zhang2023code, wan2024does}, but may have partial access to training data (e.g., from public code repositories). A common strategy in this setting is the \emph{shadow model} technique: the attacker trains one or more surrogate models on datasets with known membership labels, collects features (e.g., code representations) from these models, and then trains a binary inference classifier~\cite{yang2024gotcha, zhang2023code}. At inference time, the classifier uses analogous features extracted from the victim model to predict membership. The success of this approach hinges on how well the shadow models mimic the victim; training large shadow models is computationally expensive, and smaller surrogates may poorly approximate a large model, limiting transferability~\cite{niu2023codexleaks}.

\paragraph{Black-Box MIA}
In the black-box scenario, adversaries can only query the victim model and observe generated completions~\cite{alkaswan2024traces, yang2024unveiling}. Existing black-box MIAs employ methods that analyze completion fidelity, such as exact or fuzzy matching of predicted suffixes~\cite{alkaswan2024traces}, and statistical metrics including perplexity, perplexity ratios, and average perplexity, leveraging the observation that member examples typically yield lower perplexity and more accurate completions~\cite{yang2024unveiling}. Other approaches probe model sensitivity to small perturbations: for instance, masking individual tokens and verifying whether the masked predictions match the original tokens~\cite{majdinasab2024trained}, or comparing outputs on semantically equivalent code variants (e.g., lowercase versus uppercase) to detect disparities indicative of memorization~\cite{zhang2023code, choquette2021label}. Additionally, carefully crafted prompts can induce privacy leaks by coaxing the model to reproduce memorized code snippets, although such prompt engineering often requires manual intervention and may not generalize across different models or tasks~\cite{niu2023codexleaks}. 

AdvPrompt-MIA adopts a gray-box setting with limited training data access~\cite{yang2024gotcha}. Unlike existing gray-box methods that rely on surrogate models, our method leverages deep learning to directly capture and exploit behavioral differences under perturbations and achieve higher inference performance.

\section{Threats to Validity}

\textit{Threats to Internal Validity.}
Internal validity concerns the extent to which the study results are free from bias.
Our method leverages five predefined semantics-preserving code perturbations to induce model output variations. 
The findings may be specific to these transformations, and other perturbations could yield different results. 
To mitigate this threat, we selected several perturbations (e.g., VR and IDP) inspired by and commonly used in prior adversarial code analysis work.
Additionally, we use CodeBERT to compute similarity between perturbed and original outputs, a key step in feature vector construction.
We acknowledge that alternative code models or similarity metrics may affect the attack's effectiveness and will explore them in future work.


\textit{Threats to External Validity.}
External validity concerns the generalizability of our findings. Our evaluation focuses on widely used code models up to 7B parameters due to hardware constraints, though larger models (e.g., 34B) are increasingly adopted and remain unexplored. In addition, all experiments are conducted on unprotected models, whereas those with defense mechanisms may behave differently. Furthermore, our threat model targets the common real-world scenario of private, task-specific fine-tuning, where memorization risk is acute and existing MIAs underperform. Extending to pretraining or post-training (e.g., RLHF/RLVR) is important but non-trivial due to different training paradigms.
Future work will extend and refine AdvPrompt-MIA to apply to these scenarios.

\section{Conclusion}
In this paper, we propose AdvPrompt-MIA, a novel MIA method that leverages a series of semantics-preserving adversarial perturbations to probe the model's behavior. By extracting and learning from perturbation-induced output variations, our method achieves robust and accurate membership inference across different models and datasets. Experimental results show that AdvPrompt-MIA consistently outperforms state-of-the-art baselines and generalizes well in both cross-model and cross-dataset settings.

\section*{Acknowledgment}
We thank the anonymous reviewers for their valuable feedback. This research was supported by the National Natural Science Foundation of China under Grants No. 62302125 and 62272132, and by a Tier 1 grant funded by the Ministry of Education in Singapore (22-SIS-SMU-099 [C220/ MSS23C017]).

\bibliographystyle{IEEEtran}
\bibliography{reference}

\end{document}